\documentclass[nofootinbib,aps,superscriptaddress,preprint,floatfix]{revtex4}

\usepackage{graphicx,epsf}

\def\D0bar{\overline D{}^0}
\def\K0bar{\overline K{}^0}
\def\DDbar{D^0-\overline D{}^0}
\def\Dbar{\overline D{}^0}
\def\beq{\begin{equation}}
\def\eeq{\end{equation}}
\def\beqa{\begin{eqnarray}}
\def\eeqa{\end{eqnarray}}
\def\bea{\begin{eqnarray}}
\def\eea{\end{eqnarray}}
\def\beq{\begin{equation}}
\def\eeq{\end{equation}}

\def\Re{{\cal R \mskip-4mu \lower.1ex \hbox{\it e}\,}}
\def\Im{{\cal I \mskip-5mu \lower.1ex \hbox{\it m}\,}}
\def\be{\begin{equation}}
\def\ee{\end{equation}}

\def\Re{{\cal R \mskip-4mu \lower.1ex \hbox{\it e}\,}}
\def\Im{{\cal I \mskip-5mu \lower.1ex \hbox{\it m}\,}}

\def\sub#1{_{\lower.25ex\hbox{$\scriptstyle#1$}}}
\def\sul#1{_{\kern-.1em#1}}
\def\sll#1{_{\kern-.2em#1}}
\def\sbl#1{_{\kern-.1em\lower.25ex\hbox{$\scriptstyle#1$}}}
\def\ssb#1{_{\lower.25ex\hbox{$\scriptscriptstyle#1$}}}
\def\sbb#1{_{\lower.4ex\hbox{$\scriptstyle#1$}}}

\def\to{\rightarrow}
\def\dmix{\ifmmode D^0-\bar D^0 \else $D^0$-$\bar D^0$\fi}
\def\dm{\Delta M_D}
\def\dg{\Delta \Gamma_D}
\def\dmd{\ifmmode \Delta M_D \else $\Delta M_D$\fi}
\def\mh{\ifmmode m\sbl H \else $m\sbl H$\fi}
\def\mch{\ifmmode M_{H^\pm} \else $M_{H^\pm}$\fi}
\def\mt{\ifmmode m_t\else $m_t$\fi}
\def\mc{\ifmmode m_c\else $m_c$\fi}
\def\mz{\ifmmode M_Z\else $M_Z$\fi}
\def\mw{\ifmmode M_W\else $M_W$\fi}
\def\mws{\ifmmode M_W^2 \else $M_W^2$\fi}
\def\mhs{\ifmmode M_H^2 \else $M_H^2$\fi}
\def\mzs{\ifmmode M_Z^2 \else $M_Z^2$\fi}
\def\mts{\ifmmode m_t^2 \else $m_t^2$\fi}
\def\mcs{\ifmmode m_c^2 \else $m_c^2$\fi}
\def\mchs{\ifmmode M_{H^\pm}^2 \else $M_{H^\pm}^2$\fi}
\def\ztwo{\ifmmode Z_2\else $Z_2$\fi}
\def\zone{\ifmmode Z_1\else $Z_1$\fi}
\def\mtwo{\ifmmode M_2\else $M_2$\fi}
\def\mone{\ifmmode M_1\else $M_1$\fi}
\def\tb{\ifmmode \tan\beta \else $\tan\beta$\fi}
\def\xw{\ifmmode x\sub w\else $x\sub w$\fi}
\def\ch{\ifmmode H^\pm \else $H^\pm$\fi}
\def\lum{\ifmmode {\cal L}\else ${\cal L}$\fi}
\def\inpb{\ifmmode {\rm pb}^{-1}\else ${\rm pb}^{-1}$\fi}
\def\infb{\ifmmode {\rm fb}^{-1}\else ${\rm fb}^{-1}$\fi}
\def\epem{\ifmmode e^+e^-\else $e^+e^-$\fi}
\def\ppb{\ifmmode \bar pp\else $\bar pp$\fi}

\newskip\zatskip \zatskip=0pt plus0pt minus0pt
\def\matth{\mathsurround=0pt}

\def\gsim{\mathrel{\mathpalette\atversim>}}
\def\atversim#1#2{\lower0.7ex\vbox{\baselineskip\zatskip\lineskip\zatskip
  \lineskiplimit 0pt\ialign{$\matth#1\hfil##\hfil$\crcr#2\crcr\sim\crcr}}}

\def\Re{{\cal R \mskip-4mu \lower.1ex \hbox{\it e}\,}}
\def\Im{{\cal I \mskip-5mu \lower.1ex \hbox{\it m}\,}}

\def\sub#1{_{\lower.25ex\hbox{$\scriptstyle#1$}}}
\def\sul#1{_{\kern-.1em#1}}
\def\sll#1{_{\kern-.2em#1}}
\def\sbl#1{_{\kern-.1em\lower.25ex\hbox{$\scriptstyle#1$}}}
\def\ssb#1{_{\lower.25ex\hbox{$\scriptscriptstyle#1$}}}
\def\sbb#1{_{\lower.4ex\hbox{$\scriptstyle#1$}}}

\def\to{\rightarrow}

\def\rb{\ifmmode R_b\else $R_b$\fi}
\def\rc{\ifmmode R_c\else $R_c$\fi}
\def\ac{\ifmmode A_c\else $A_c$\fi}
\def\dmix{\ifmmode D^0-\bar D^0 \else $D^0$-$\bar D^0$\fi}
\def\dm{\ifmmode \Delta M_D \else $\Delta M_D$\fi}
\def\rb{\ifmmode R_b\else $R_b$\fi}
\def\mh{\ifmmode m\sbl H \else $m\sbl H$\fi}
\def\mch{\ifmmode M_{H^\pm} \else $M_{H^\pm}$\fi}
\def\mt{\ifmmode m_t\else $m_t$\fi}
\def\mc{\ifmmode m_c\else $m_c$\fi}
\def\mz{\ifmmode M_Z\else $M_Z$\fi}
\def\mw{\ifmmode M_W\else $M_W$\fi}
\def\mws{\ifmmode M_W^2 \else $M_W^2$\fi}

\def\mhs{\ifmmode m_H^2 \else $m_H^2$\fi}
\def\mzs{\ifmmode M_Z^2 \else $M_Z^2$\fi}
\def\mts{\ifmmode m_t^2 \else $m_t^2$\fi}
\def\mcs{\ifmmode m_c^2 \else $m_c^2$\fi}
\def\mchs{\ifmmode m_{H^\pm}^2 \else $m_{H^\pm}^2$\fi}
\def\ztwo{\ifmmode Z_2\else $Z_2$\fi}
\def\zone{\ifmmode Z_1\else $Z_1$\fi}
\def\mtwo{\ifmmode M_2\else $M_2$\fi}
\def\mone{\ifmmode M_1\else $M_1$\fi}
\def\bsg{\ifmmode b\to s\gamma\else $b\to s\gamma$\fi}
\def\tb{\ifmmode \tan\beta \else $\tan\beta$\fi}
\def\xw{\ifmmode x\sub w\else $x\sub w$\fi}
\def\ch{\ifmmode H^\pm \else $H^\pm$\fi}
\def\lum{\ifmmode {\cal L}\else ${\cal L}$\fi}
\def\inpb{\ifmmode {\rm pb}^{-1}\else ${\rm pb}^{-1}$\fi}
\def\infb{\ifmmode {\rm fb}^{-1}\else ${\rm fb}^{-1}$\fi}
\def\epem{\ifmmode e^+e^-\else $e^+e^-$\fi}
\def\ppb{\ifmmode \bar pp\else $\bar pp$\fi}

\def\be{\begin{equation}}
\def\ee{\end{equation}}

\begin{document}
\vspace{3.0cm}
\preprint{\vbox 
{
\hbox{SLAC-PUB-13565}
\hbox{WSU--HEP--0901} \hbox{UH-511-1136-09}
}}

\vspace*{2cm}

\title{\boldmath 
Relating $D^0$-${\bar D}^0$ Mixing and $D^0\to \ell^+\ell^-$ 
with New Physics}

\author{Eugene Golowich}
\affiliation{Department of Physics,
        University of Massachusetts\\[-6pt]
        Amherst, MA 01003}

\author{JoAnne Hewett}
\affiliation{SLAC National Accelerator Laboratory, 2575 Sand Hill Rd, 
Menlo Park, CA, 94025, USA}

\author{Sandip Pakvasa}
\affiliation{Department of Physics and Astronomy\\[-6pt]
        University of Hawaii, 
        Honolulu, HI 96822}

\author{Alexey A.\ Petrov\vspace{8pt}}
\affiliation{Department of Physics and Astronomy\\[-6pt]
        Wayne State University, Detroit, MI 48201}

\affiliation{Michigan Center for Theoretical Physics\\[-6pt]
        University of Michigan, Ann Arbor, MI 48196\\[-6pt] $\phantom{}$ }

\begin{abstract}
We point out how, in certain models of New Physics, the same
combination of 
couplings occurs in the amplitudes for both $D^0$-${\bar D}^0$ mixing
and the rare decays $D^0\to \ell^+\ell^-$.  If the New Physics 
dominates and is responsible for the observed mixing, then a very simple
correlation exists between the magnitudes of each; in fact the rates for
the decay $D^0\to \ell^+\ell^-$ are completely fixed by the mixing.
Observation of $D^0\to\ell^+\ell^-$ in excess of the Standard Model
prediction could identify New Physics contributions to $D^0$-${\bar D}^0$
mixing.
\vskip 1in
%
\end{abstract}

\def\thepage{{}}
\maketitle
\def\thepage{\arabic{page}}

\section{Introduction}
Following many years of effort, there is now indisputable 
experimental evidence for $D^0$-${\bar D}^0$ mixing.  
The current values (the HFAG `no CPV-allowed' fit~\cite{hfag}) 
of the $D^0$ mixing parameters are 
\begin{eqnarray}
& & x_{\rm D} \equiv {\Delta M_{\rm D} \over \Gamma_{\rm D}} = 
0.0100^{+0.0024}_{-0.0026} \qquad \text{and} \qquad 
y_{\rm D} \equiv {\Delta \Gamma_{\rm D} \over 2 \Gamma_{\rm D}} =
0.0076^{+0.0017}_{-0.0018}  \ \ .
\label{dmix1}
\end{eqnarray}
These show that (i) charm mixing occurs
at about the percent level, (ii) $x_{\rm D},y_{\rm D}$
are comparable in magnitude and (iii) the signs of $x_{\rm D}$
and $y_{\rm D}$ are positive (although a direct measurement of
the sign of $x_{\rm D}$ is yet to be made).

While it is quite likely that the observed mixing amplitude is 
dominated by the Standard Model contributions, the exact predictions 
are quite difficult.\footnote{Henceforth, we will make frequent use of the 
abbreviations SM for Standard Model and NP for New Physics.} 
  There are several reasons for this~\cite{Artuso:2008vf,proba,probb}. 
For example, in the ``short distance'' approach~\cite{Georgi:1992as} at 
leading order in the Operator Product Expansion (OPE) formalism (operators 
of dimension $D=6$), the individual diagrams are CKM-suppressed to the level 
${\cal O}(\lambda^2)$ ($\lambda \simeq 0.22$ is the familiar Wolfenstein parameter), 
hinting that the observed charm mixing is a simple consequence of 
CKM structure.  This is, however, not correct because severe 
cancellations between diagrams (even through ${\cal O}(\alpha_s)$) 
greatly reduce the $D=6$ mixing to 
${\cal O}(10^{-6})$~\cite{proba,probb}.  
As for higher ($D>6$) orders in OPE, it is true that 
certain enhanced contributions have been 
identified~\cite{Ohl:1992sr,Bigi:2000wn}, but a definitive evaluation 
is lacking due to the large 
number of $D>6$ operators and the inability to determine their matrix 
elements.  A promising alternative approach which involves a 
hadron-level description~\cite{Falk:2004wg} 
may be able to account for the observed 
magnitude of $y_{\rm D}$ and $x_{\rm D}$, 
but  predicts their relative sign to be opposite. It is fair to say that
this is probably not the final word on the SM analysis.

Given the uncertain status of the SM description, 
it would be tempting but premature~\cite{Nir:2007ac} to attribute the 
observed $x_{\rm D}$ to New Physics.\footnote{We will focus on $x_{\rm D}$ 
in this paper.  Not only does the SM estimate for 
$y_{\rm D}$ work reasonably well when long distance effects are
included~\cite{Falk:2001hx}, but it has also been shown that NP
effects are too small to have any significant 
impact~\cite{Golowich:2006gq,Yeghiyan:2007uj,Petrov:2007gp}.}   
But clearly, the possibility that NP makes a 
significant or even dominant contribution to
the observed mixing is open. A recent comprehensive treatment of NP 
models~\cite{Golowich:2007ka} shows that a large number of such models 
can accommodate a value of $x_{\rm D}$ at the per cent level.  
This encourages us to further explore the NP option.  
In particular, New Physics could affect charm-related 
processes beyond mixing, such as rare decays~\cite{Fajfer:2007dy}. 
Of special interest are the $D^0 \to \ell^+ \ell^-$ decays.  
At present, there are only the upper 
limits~\cite{PDG,Aubert:2004bs,Abt:2004hn,Acosta:2003ag} 
\beqa
{\cal B}_{D^0 \to \mu^+\mu^-} &\le & 1.3\times 10^{-6}, \qquad 
{\cal B}_{D^0 \to e^+ e^-} \le 1.2\times 10^{-6}, \qquad  {\rm and} 
\nonumber \\
{\cal B}_{D^0 \to \mu^\pm e^\mp} &\le & 8.1\times 10^{-7} \ \ ,
\label{brs}
\eeqa
all at CL=$90\%$. Such branching fractions place bounds
on possible NP couplings, which can be compared
with that obtained from $D^0$-${\bar D}^0$
mixing.  In this paper we study the impact of NP 
on the combined system of $D^0$-${\bar D}^0$ mixing and
the rare decays $D^0\to \ell^+\ell^-$.  It should be stressed that
the SM rate for the decay mode, ${\cal B}_{D^0 \to \mu^+\mu^-} 
\simeq 3 \times 10^{-13}$ can be estimated fairly reliably 
even upon accounting for the effect of long distance 
enhancement~\cite{Burdman:2001tf}. This smallness of the SM signal 
makes it easier for NP contributions to stand out. In this paper, 
we point out how, in certain NP models, the 
{\it same} couplings occur in the amplitudes for both 
$D^0$-${\bar D}^0$ mixing and $D^0\to \ell^+\ell^-$ decay.
If the NP effects are significant in mixing, then a correlation will exist 
between the magnitudes of each. In fact the correlation can be very simple
and striking, with the branching fraction ${\cal B}_{D^0 \to \ell^+\ell^-}$  
being proportional to the mixing parameter $x_{\rm D}$.

For each NP model considered in this paper, we shall make the simplifying 
assumption that the NP dominates $D^0$-${\bar D}^0$ mixing and then 
derive the correlated branching fraction that is then predicted. 
Obviously, if NP does {\it not} dominate, all
our results for the branching fractions become upper bounds.  
We shall stress the general issue of which conditions 
allow for such correlations and give specific examples. 
Finally, even if the number of parameters in a given NP model 
is too large to give a unique prediction for 
${\cal B}_{D^0 \to \ell^+\ell^-}$ in terms of $x_{\rm D}$ 
({\it e.g.} $Z'$ models, {\it etc}), we show 
in Sects.~III~B-D how it is possible to 
estimate the scale of ${\cal B}_{D^0 \to \ell^+\ell^-}$ by using 
the value of $x_{\rm D}$ as input.

\section{Effective Lagrangians} 

Heavy particles present in NP models are not produced in 
final states of charm quark decays. Yet, effects generated by 
exchanges of these new particles can be accounted for in 
effective operators built out of the SM degrees of freedom. 
That is, by integrating out degrees of freedom associated with 
new interactions at a heavy scale $M$, we obtain 
an effective hamiltonian written in the form of a series of operators of 
increasing dimension. Here, we restrict our attention to the 
leading order operators, of dimension $D=6$.  
For both $\DDbar$ mixing and $D^0\to \ell^+\ell^-$ decays, 
the complete basis of effective operators is known and is 
expressed most conveniently in terms of chiral quark fields,
\beq\label{SeriesOfOperators}
\langle f | {\cal H}_{NP} | i \rangle =
G \sum_{i=1} {\rm C}_i (\mu) ~
\langle f | Q_i  | i \rangle (\mu) \ \ ,
\eeq
where the prefactor $G$ has the dimension of inverse-squared mass, the 
${\rm C}_i$ are dimensionless Wilson coefficients, and the $Q_i$ are 
the effective operators of dimension six.  Throughout, our convention for 
defining chiral projections for a field $q(x)$ will be 
$q_{L,R}(x) \equiv(1 \pm \gamma_5) q(x)/2$.

For $\Delta C = 2$ processes, there are eight effective operators 
that can contribute~\cite{Ciuchini:1997bw,Golowich:2007ka},
\beqa
\begin{array}{l}
Q_1 = (\overline{u}_L \gamma_\mu c_L) \ (\overline{u}_L \gamma^\mu
c_L)\ , \\
Q_2 = (\overline{u}_L \gamma_\mu c_L) \ (\overline{u}_R \gamma^\mu
c_R)\ , \\
Q_3 = (\overline{u}_L c_R) \ (\overline{u}_R c_L) \ , \\
Q_4 = (\overline{u}_R c_L) \ (\overline{u}_R c_L) \ ,
\end{array}
\qquad
\begin{array}{l}
Q_5 = (\overline{u}_R \sigma_{\mu\nu} c_L) \ ( \overline{u}_R
\sigma^{\mu\nu} c_L)\ , \\
Q_6 = (\overline{u}_R \gamma_\mu c_R) \ (\overline{u}_R \gamma^\mu
c_R)\ , \\
Q_7 = (\overline{u}_L c_R) \ (\overline{u}_L c_R) \ , \\
Q_8 = (\overline{u}_L \sigma_{\mu\nu} c_R) \ (\overline{u}_L
\sigma^{\mu\nu} c_R)\ \ .
\end{array}
\label{SetOfOperators}
\eeqa
These operators are generated 
at the scale $M$ where the NP is integrated out. A non-trivial 
operator mixing then occurs via renormalization group 
running of these operators between the heavy scale $M$ and the light 
scale $\mu$ at which hadronic matrix elements are computed.

All possible NP contributions to $c \to u \ell^+ \ell^-$ 
can be similarly summarized.  In this case, however, there are now 
ten operators, 
\beqa
\begin{array}{l}
\widetilde Q_1 = (\overline{\ell}_L \gamma_\mu \ell_L) \ 
(\overline{u}_L \gamma^\mu
c_L)\ , \\
\widetilde Q_2 = (\overline{\ell}_L \gamma_\mu \ell_L) \ 
(\overline{u}_R \gamma^\mu
c_R)\ , \\ 
\widetilde Q_3 = (\overline{\ell}_L \ell_R) \ (\overline{u}_R c_L) \ , 
\end{array}
\qquad 
\begin{array}{l}
\widetilde Q_4 = (\overline{\ell}_R \ell_L) \ 
(\overline{u}_R c_L) \ , \\
\widetilde Q_5 = (\overline{\ell}_R \sigma_{\mu\nu} \ell_L) \ 
( \overline{u}_R \sigma^{\mu\nu} c_L)\ ,\\
\phantom{xxxxx} 
\end{array}
\label{SetOfOperatorsLL}
\eeqa
with five additional operators $\widetilde Q_6, \dots, \widetilde Q_{10}$ 
being obtained respectively from those in Eq.~(\ref{SetOfOperatorsLL}) by 
the substitutions $L \to R$ and $R \to L$.
The corresponding Wilson coefficients will be denoted as 
$\widetilde C_i(\mu)$. 
It is worth noting that only eight operators contribute to
$D^0\to \ell^+\ell^-$, as 
$\langle \ell^+ \ell^- | \widetilde Q_5 | D^0 \rangle =
\langle \ell^+ \ell^- | \widetilde Q_{10} | D^0 \rangle = 0$. 

To obtain a general expression for $x_{\rm D}$ as implied by the 
effective Hamiltonian of Eq.~(\ref{SeriesOfOperators}), we 
evaluate the $D^0$-to-${\bar D}^0$ matrix element in the modified vacuum
saturation approximation of Appendix~A and work at the light 
scale $\mu = m_c$, 
\begin{eqnarray}\label{xDgen}
& & x_{\rm D} = G {f_D^2 M_D B_D \over
\Gamma_D} \bigg[ {2 \over 3}(C_1(m_c) + C_6(m_c)) - 
\left[\displaystyle{1 \over 2} + \displaystyle{\eta \over 3} \right]
C_2(m_c) + \left[ \displaystyle{1 \over 12} + 
\displaystyle{\eta \over 2} \right]
C_3(m_c) \nonumber \\
& & \hspace{3.3cm} - {5 \eta \over 12}(C_4(m_c) + C_7(m_c))
 + \eta (C_5(m_c) + C_8(m_c)) \bigg] 
\ \ ,
\end{eqnarray}
where we have taken $N_c = 3$, we remind the reader that $\eta$ 
is discussed in Appendix~A and the 
prefactor $G$ defines the scale at which NP is integrated out. 
To use this expression one must relate the light-scale coefficients 
$\{ C_i(m_c)\}$ to their heavy-scale counterparts $\{C_i(M)\}$ 
in terms of the RG-running factors given in Appendix~A.

The rare decays $D^0\to \ell^+\ell^-$ and 
$D^0\to \mu^+e^-$ are treated analogously. 
To the decay amplitude 
\beq\label{decayampl}
{\cal M} = {\bar u}({\bf p}_-, s_-) \left[ A + B \gamma_5 
\right] v({\bf p}_+, s_+) \ \ , 
\eeq
are associated the branching fractions 
\begin{eqnarray}\label{Dllgen}
& & {\cal B}_{D^0 \to \ell^+\ell^-} = 
\frac{M_D}{8 \pi \Gamma_{\rm D}} \sqrt{1-\frac{4 m_\ell^2}{M_D^2}}
\left[ \left(1-\frac{4 m_\ell^2}{M_D^2}\right)\left|A\right|^2  +
\left|B\right|^2 \right] \ \ , 
\nonumber \\
& & {\cal B}_{D^0 \to \mu^+e^-} = 
\frac{M_D}{8 \pi \Gamma_{\rm D}} 
\left( 1-\frac{ m_\mu^2}{M_D^2} \right)^2 
\left[ \left|A\right|^2  + \left|B\right|^2 \right] \ \ ,
\end{eqnarray}
where electron mass has been neglected in the latter expression.
Any NP contribution described by the operators of
Eq.~(\ref{SetOfOperatorsLL}) gives for 
the amplitudes $A$ and $B$, 
\begin{eqnarray}
\left| A\right|  &=& G \frac{f_D M_D^2}{4 m_c} \left[\widetilde C_{3-8} + 
\widetilde C_{4-9}\right]\ , \nonumber 
 \\
\left| B\right|  &=& G \frac{f_D}{4} \left[
2 m_\ell \left(\widetilde C_{1-2} + \widetilde C_{6-7}\right)
+  \frac{M_D^2}{m_c}
\left(\widetilde C_{4-3} + \widetilde C_{9-8}\right)
\right]\ ,  \label{DlCoeff}
\end{eqnarray}
with $\widetilde C_{i-k} \equiv \widetilde C_i-\widetilde C_k$. 
In general, one cannot predict 
the rare decay rate by knowing just the mixing rate, even if
both $x_D$ and ${\cal B}_{D^0 \to \ell^+\ell^-}$ are dominated by a 
given NP contribution. We shall see, however, that this is 
possible for a restricted subset of NP models.

\section{NP Models with tree-level amplitudes} 

This is the most obvious situation for producing a correlation 
between mixing and decay because there is a factorization 
between the initial and final interaction vertices.  In the following, 
it will be convenient to consider separately the propagation 
of a spin-1 boson V and of a spin-0 boson S 
as the intermediate particle in the tree-level amplitudes.  
The bosons V and S can be of either parity.

\vspace{0.3cm}

{\it Spin-1 Boson V}: 
Assuming that the spin-1 
particle $V$ has flavor-changing 
couplings and keeping all the operators in the effective 
Lagrangian up to dimension 5, 
the most general Lagrangian can be written as
\beq
{\cal H}_V = {\cal H}^{\rm FCNC}_V+ {\cal H}^L_V \ \ ,
\eeq
where the quark part ${\cal H}^{\rm FCNC}_V$ is 
\bea\label{VecHam}
{\cal H}^{\rm FCNC}_V = g_{V1} \overline u_L \gamma_\mu c_L V^\mu + 
g_{V2} \overline u_R \gamma_\mu c_R V^\mu + 
g_{V3} \overline u_L \sigma_{\mu\nu} c_R V^{\mu\nu} + 
g_{V4} \overline u_R \sigma_{\mu\nu} c_L V^{\mu\nu} 
\eea
and the part that describes interactions of 
$V$ with leptons ${\cal H}^L_V$ is 
\bea\label{VecHamLept}
{\cal H}^L_V = g_{V1}^\prime \overline \ell_L \gamma_\mu \ell_L V^\mu + 
g_{V2}^\prime \overline \ell_R \gamma_\mu \ell_R V^\mu + 
g_{V3}^\prime \overline \ell_L \sigma_{\mu\nu} \ell_R V^{\mu\nu} + 
g_{V4}^\prime \overline \ell_R \sigma_{\mu\nu} \ell_L V^{\mu\nu} .
\eea
Here $V_\mu$ is the vector field 
and $V_{\mu\nu}=\partial_\mu V_\nu - \partial_\nu V_\mu + ...$ 
is the field-strength tensor for $V_\mu$. For this study it is 
not important whether the field 
$V$ corresponds to an abelian or non-abelian gauge symmetry group.

In order to see the leading contribution to $D$ mixing from Eq.~(\ref{VecHam}), 
let us consider a correlator,
\beq\label{Correlator}
\Sigma_D (q^2) = (-i) \int d^4 x ~e^{i(q-p)\cdot x} \langle \Dbar (p) | 
T \left\{{\cal H}^{\Delta C=1} (x) 
{\cal H}^{\Delta C=1} (0) \right\} | D^0 (p) \rangle.
\eeq
This correlator is related to the mass and lifetime
differences of a $D$-meson as~\cite{Falk:2004wg},
\beq
\Sigma_D (M_D^2) = 2 M_D \left(\dm - \frac{i}{2} \dg \right).
\eeq
Inserting Eq.~(\ref{VecHam}) for ${\cal H}^{\Delta C=1}$, we obtain 
\bea
\Sigma_D (q^2) &=& (-i) \ \beta \int d^4 x ~ e^{i(q-p)\cdot x}  
\langle 0 |T\left\{ V^\mu (x) V^\nu (0) \right\} | 0 \rangle
\langle \Dbar (p)| \
g_{V1}^2 \ \overline{u}_L \gamma_\mu c_L (x) \ 
\overline{u}_L \gamma_\nu c_L (0) 
\nonumber \\
&+& g_{V1} g_{V2} \ \overline{u}_L \gamma_\mu c_L (x) 
\ \overline{u}_R \gamma_\nu c_R (0)
+ g_{V1} g_{V2} \ \overline{u}_R \gamma_\mu c_R (x) 
\ \overline{u}_L \gamma_\nu c_L (0) \nonumber 
\\
&+& g_{V2}^2 \ \overline{u}_R \gamma_\mu c_R (x) 
\ \overline{u}_R \gamma_\nu c_R (0)
\ + \ {\cal O}(1/M_V) \ | D^0 (p)\rangle \ , 
\label{selfen}
\eea
where ${\cal O}(1/M_V)$ denotes terms additionally 
suppressed by powers of 
$1/M_V$ and $\beta \sim {\cal O}(1)$ denotes all 
relevant traces over group indices 
associated with $V_\mu$, with $\beta=1$ for an Abelian symmetry group.
The leading-order $D=6$ contribution is found by expanding the
vector boson propagator in the large $M_V$-limit and then performing 
the resulting elementary integral,  
\bea
\Sigma_D (M_D^2) = \frac{\beta}{M_V^2} ~
\langle \Dbar (p) | \ 
g_{V1}^2 Q_1 + 2 g_{V1} g_{V2} Q_2 + g_{V2}^2 Q_6 
\ | D^0 (p)\rangle \ .
\eea
Taking into account RG-running between the heavy scale $M_V$ and
the light scale $\mu = m_c$ at which the matrix elements are computed, 
we obtain a subcase of the general Eq.~(\ref{xDgen}), 
\bea\label{dMV2}
x_{\rm D}^{\rm (V)} &=& {\beta f_D^2 M_D B_D \over 2 M_{V}^2 \Gamma_D} 
 \left[ 
{2 \over 3} \left( C_1 (m_c) + C_6 (m_c) \right) - 
\left[\displaystyle{1 \over 2} + \displaystyle{\eta \over 3} \right]
C_2(m_c) + \left[ \displaystyle{1 \over 12} + 
\displaystyle{\eta \over 2} \right] C_3(m_c) \right]  
\eea
where the superscript on $x_{\rm D}^{\rm (V)}$ denotes propagation of 
a vector boson in the tree amplitude.  
The Wilson coefficients evaluated at scale $\mu = m_c$ are 
\bea
\begin{array}{l}
{\rm C}_1(m_c) = r (m_c,M_V)~ g_{V1}^2\ , \nonumber \\
{\rm C}_2(m_c)= 2 \ r(m_c,M_V)^{1/2} g_{V1} g_{V2}   \ , 
\end{array}
\quad 
\begin{array}{l}
{\rm C}_3(m_c)= \frac{4}{3} \left[
r(m_c,M_V)^{1/2} - r(m_c,M_V)^{-4}
\right] g_{V1} g_{V2} \ , \nonumber \\
{\rm C}_6(m_c)= r (m_c,M_V) ~ g_{V2}^2 \ \ . 
\end{array} \\
\label{zwilsons}
\eea

Similar calculations can be performed for the 
$D^0 \to \ell^+\ell^-$ decay.  The effective Hamiltonian 
in this case is 
\beq
{\cal H}_{c\to u\ell^+\ell^-}^{\rm (V)} = \frac{1}{M_V^2} 
\left[
g_{V1}g_{V1}' \widetilde Q_1 +
g_{V1}g_{V2}' \widetilde Q_7 + g_{V1}'g_{V2}  \widetilde Q_2 +
g_{V2}g_{V2}'  \widetilde Q_6 
\right]\ ,
\eeq
which leads to the branching fraction, 
\beq\label{GammaV}
{\cal B}_{D^0 \to \ell^+\ell^-}^{\rm (V)} = \frac{f_D^2 m_\ell^2 M_D}
{32\pi M_V^4 \Gamma_D}
\sqrt{1-\frac{4 m_\ell^2}{M_D^2}} 
\left(g_{V1} - g_{V2}\right)^2 \left(g_{V1}' - g_{V2}'\right)^2\ .
\eeq
Clearly, Eqs.~(\ref{dMV2}) and (\ref{GammaV}) can be related 
to each other only for a specific set of NP models. We shall consider 
those shortly. 

\vspace{0.3cm}

{\it Spin-0 Boson S}: 
Analogous procedures can be followed if now the FCNC is generated 
by quarks interacting with spin-0 particles. Again, 
assuming that the spin-0 particle $S$ has flavor-changing couplings 
and keeping all the operators in the effective Hamiltonian up to 
dimension five, we can write the most general Hamiltonian as
\beq
{\cal H}_S = {\cal H}^{\rm FCNC}_S + {\cal H}^L_S \ \ ,
\eeq
where the quark FCNC part is given by
\bea\label{SFCNC}
{\cal H}^{\rm FCNC}_S = g_{S 1} \overline u_L c_R S + 
g_{S 2} \overline u_R c_L S + 
g_{S 3} \overline u_L \gamma_\mu c_L \partial^\mu S + 
g_{S 4} \overline u_R \gamma_\mu c_R \partial^\mu S 
\eea
and the part that is responsible for the interactions of $S$ 
with leptons is 
\bea
{\cal H}^L_S = g_{S 1}^\prime  \overline \ell_L \ell_R S + 
g_{S 2}^\prime  \overline \ell_R \ell_L S + 
g_{S 3}^\prime \overline \ell_L \gamma_\mu \ell_L \partial^\mu S + 
g_{S 4}^\prime \overline \ell_R \gamma_\mu \ell_R \partial^\mu S \ \ .
\eea
Inserting this Hamiltonian into the correlator Eq.~(\ref{Correlator}) 
and performing steps similar to the spin-one case leads to 
\bea\label{dMS}
\Sigma_D (M_D^2) &=& - \frac{1}{M_S^2} \langle \Dbar (p) | \ 
g_{S 1}^2 Q_7 + 
2 g_{S 1} g_{S 2} Q_3 + g_{S 2}^2 Q_4 \ | D^0 (p) \ \ .
\eea
%
Evaluation at scale $\mu = m_c$ gives 
\bea
x_D^{\rm (S)} = - \frac{f_D^2 M_D B_D}{2 \Gamma_D M_S^2}
\left[ \left[ \displaystyle{1 \over 12} + 
\displaystyle{\eta \over 2} \right]C_3(m_c) - 
{5 \eta \over 12}(C_4(m_c) + C_7(m_c))
+ \eta (C_5(m_c)+C_8(m_c) )
\right] \nonumber \\
\label{dMS2}
\eea
with the Wilson coefficients defined as
\bea
& & C_3(m_c) = - 2 r(m_c,M_S )^{-4} ~g_{S 1} g_{S 2}
\nonumber \\
& & C_4(m_c) = - \left[
\left(\frac{1}{2}-\frac{8}{\sqrt{241}} \right) 
r_+ (m_c,M_S) +  
\left(\frac{1}{2}+\frac{8}{\sqrt{241}} \right) 
r_- (m_c,M_S) \right] g_{S 2}^2 
\nonumber \\
& & C_5(m_c) = -\frac{1}{8 \sqrt{241}} \left[
r_+ (m_c,M_S) - r_- (m_c,M_S)\right] g_{S 2}^2
\label{cs1} \\
& & C_7(m_c)= - \left[
\left(\frac{1}{2}-\frac{8}{\sqrt{241}} \right) 
r_+ (m_c,M_S) +  
\left(\frac{1}{2}+\frac{8}{\sqrt{241}} \right) 
r_- (m_c,M_S) \right] g_{S 1}^2 
\nonumber \\
& & C_8(m_c) = - \frac{1}{8 \sqrt{241}} \left[
r_+ (m_c,M_S) - r_- (m_c,M_S) \right] g_{S 1}^2 \ \ , 
\nonumber 
\eea
where for notational simplicity we have defined 
$r_\pm \equiv r^{(1 \pm \sqrt{241})/6}$  ({\it cf} Eq.~(A1)).

The effective Hamiltonian for the $D^0 \to \ell^+\ell^-$ decay 
is 
\beq
{\cal H}_{c\to u\ell^+\ell^-}^{\rm (S)} = - \frac{1}{M_S^2} 
\left[
g_{S1}g_{S1}' \widetilde Q_9 +
g_{S1}g_{S2}' \widetilde Q_8 + g_{S1}'g_{S2}  \widetilde Q_3 +
g_{S2}g_{S2}'  \widetilde Q_4 
\right]\ ,
\eeq
and from this, it follows that the branching fraction is 
\bea\label{GammaS}
& & {\cal B}_{D^0 \to \ell^+\ell^-}^{\rm (S)} 
= \frac{f_D^2 M_D^5}
{128\pi m_c^2 M_S^4 \Gamma_D}
\sqrt{1-\frac{4 m_\ell^2}{M_D^2}} 
\left(g_{S 1} - g_{S 2}\right)^2 
\left[\left(g_{S 1}' + g_{S 2}'\right)^2 
\left(1-\frac{4 m_\ell^2}{M_D^2}\right) + 
\left(g_{S 1}' - g_{S 2}'\right)^2
\right]. \nonumber \\
\eea
Note that if the spin-0 particle $S$ only has {\it scalar} FCNC couplings, 
{\it i.e.} $g_{S 1} = g_{S 2}$, no contribution to $D^0 \to \ell^+\ell^-$
branching ratio is generated at the tree level; the non-zero 
contribution to rare decays is produced at one-loop level. This follows from 
the {\it pseudoscalar} nature of the $D$-meson.

If there are not one but several particles mediating those 
processes (assuming that they all couple to quarks and leptons), the 
above generic Lagrangians would need to be modified. For example, 
in the spin-0 case, one would have to replace $g_i ~S$ with 
$ \sum_k g_{ik} S_k F(k)$, where $F(k)$ is a 
numerical factor, as $S_k$ are the mediating fields. For example, in 
models with extra dimensions the factor $F(k)$ would be related to 
Kaluza-Klein decompositions of bosons living in the bulk. 
Similar corrections have to be performed in the case of a bulk spin-1 
boson.

Below, we consider generic models where the correlations 
between the $\DDbar$ mixing rates and $D^0 \to \ell^+\ell^-$ 
rare decays can be found.

\subsection{Heavy Vector-like Quarks: Q = +2/3 Singlet Quark}

Scenarios with heavy quarks beyond the three generations are 
severely constrained experimentally, if those quarks have chiral couplings.
We thus examine the case where the heavy quarks are $SU(2)_L$ singlets 
(so-called {\it vector-like} quarks)~\cite{branco}. Here, we consider 
the charge assignment $Q = +2/3$ for the heavy quark 
and then $Q = -1/3$ in the next Section.  Weak isosinglets with $Q = +2/3$ 
occur in Little Higgs theories~\cite{q3a,q3b} in which the 
Standard Model Higgs boson is a pseudo-Goldstone boson, and the heavy
iso-singlet $T$ quark cancels the quadratic divergences generated
by the top quark when performing quantum corrections to the mass of the 
Higgs boson. Weak isosinglets with $Q = -1/3$ appear in
$E_6$ GUTs~\cite{q2,jlhtgr}, with one for each of the three
generations ($D$, $S$, and $B$).  

The presence of such quarks violates the Glashow-Weinberg-Paschos naturalness
conditions for neutral currents~\cite{gw77}.  Since their 
electroweak quantum number assignments are different than those for the
SM fermions, flavor changing neutral current interactions are generated
in the left-handed up-quark sector.  Thus, in addition to the charged current 
interaction 
\beq
{\cal L}_{\rm int}^{(ch)} = {g \over \sqrt{2}}~V_{\alpha i}
{\bar u}_{\alpha,L} \gamma_\mu d_{i,L} W^\mu\ \ , 
\eeq
there are also FCNC couplings with the $Z^0$ boson~\cite{branco}, 
\beq
{\cal L}_{\rm int}^{(ntl)} = {g \over 2\sqrt{2}\cos\theta_w}~\lambda_{ij} 
{\bar u}_{i,L} \gamma_\mu u_{j,L} Z^{0\mu}\ \ . 
\eeq
Here, $g$ is the SM $SU(2)$ gauge coupling and 
$V_{\alpha i}$ is a $4\times 3$ mixing matrix with $\alpha$ running
over $1\to 4$, $i=1\to 3$, with the CKM matrix comprising the first 
$3\times 3$ block.

\vspace{0.2cm} 

\begin{figure} [t]
\centerline{
\includegraphics[width=10cm,angle=0]{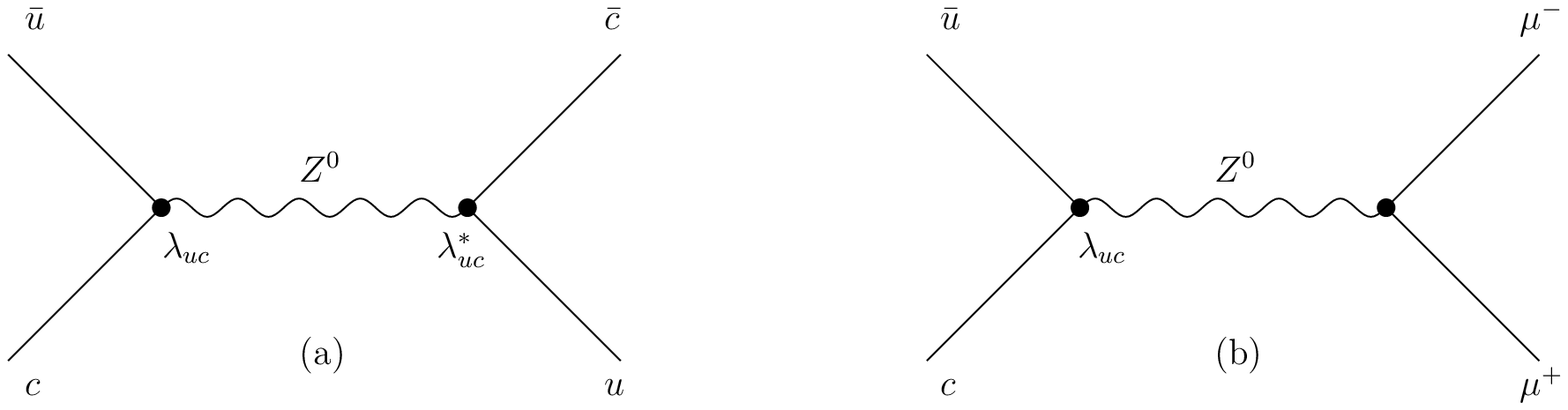}}
\caption{(a) $D^0$-${\bar D}^0$ Mixing, (b) $D^0\to \mu^+\mu^-$. 
\label{qfig2}}
\end{figure}

\noindent{\it $D^0$-${\bar D}^0$ Mixing}: 
In this case, a tree-level contribution to $\Delta M_D$ is 
generated from $Z^0$-exchange as shown in Fig.~\ref{qfig2}.  
This is represented by an effective hamiltonian at the scale $M_Z$ as 
\beq
{\cal H}_{2/3} = {g^2 \over 8 \cos^2 \theta_w M_Z^2} 
\lambda_{uc}^2 ~ Q_1 \ = \ {G_F \lambda_{uc}^2 \over \sqrt{2}} Q_1 
\ \ , 
\eeq
where from unitarity,  
\beq
\lambda_{uc} \equiv - \left( V_{ud}^*V_{cd} + V_{us}^*V_{cs} + 
V_{ub}^*V_{cb} \right) \ \ .
\eeq
Thus, we find 
\beq
x_{\rm D}^{\rm (+2/3)} = {2 G_F \lambda_{uc}^2 f_D^2 
M_D B_D r (m_c,M_Z) \over 3 \sqrt{2} \Gamma_D} 
\eeq
Using $r (m_c,M_Z) = 0.778$ and demanding that the 
NP contribution is responsible for 
the observed mixing value yields ${\lambda_{uc} = 2.39\times 10^{-4}}$. 

\vspace{0.2cm} 

\noindent {\it $D^0 \to \mu^+\mu^-$ Decay}: For this case, 
the leptons have SM couplings to the $Z^0$.  We then have 
\beqa
& & A_{D^0\to \ell^+\ell^-} = 0 \qquad B_{D^0\to \ell^+\ell^-} = 
 \lambda_{uc} {G_F f_{\rm D} m_\mu \over 2} \ \ .
\eeqa
Restricting our attention to the $\mu^+\mu^-$ final state 
because the decay amplitude is proportional to lepton mass, 
we find 
\beqa 
& & \lambda_{uc}^2 \le 8 \pi {\cal B}_{D^0 \to \mu^+\mu^-}
{\Gamma_{\rm D} \over  M_{\rm D}}  \left( {2 \over G_F f_{\rm D} m_\mu}
\right)^2 \left[ 1 - {4 m_\mu^2 \over M_{\rm D}} \right]^{-1/2} \ .
\label{x4}
\eeqa
From the branching fraction bound of Eq.~(\ref{brs}), 
we obtain ${\lambda_{uc}   \le 4.17\times 10^{-2}}$, which is 
much less restrictive than the value from $D^0$ mixing. 

\vspace{0.2cm} 

\noindent {\it Combining the Mixing and Decay Relations}: 
A correlation will exist in this case because 
the coupling between $Z^0$ and the lepton pair is known from 
the Standard Model. Thus, 
if we assume that all the $D$ meson mixing 
comes from the $Q=+2/3$ heavy quark 
({\it i.e.} $x_{\rm D}^{\rm (+2/3)} = x_{\rm D}$), then we can 
remove the dependence on the NP parameter $\lambda_{uc}$ and 
predict ${\cal B}_{D^0 \to \mu^+\mu^-}$ in terms of $x_{\rm D}$, 
\beqa
{\cal B}_{D^0 \to \mu^+\mu^-} &=&  {3 \sqrt{2} \over 64 \pi} 
~{G_F m_\mu^2 x_{\rm D}\over B_{\rm D} r(m_c,M_Z)} 
\left[ 1 - {4 m_\mu^2 \over M_{\rm D}} \right]^{1/2} 
\ \   \nonumber \\
& \simeq & \  4.3\times 10^{-9} x_{\rm D}\ \le \ 4.3\times 10^{-11} 
\  \ . \label{x5}
\eeqa

\subsection{New Gauge Boson $Z'$}

New heavy neutral gauge bosons can exist in a 
variety of NP models~\cite{Langacker:2008yv}. 
In these scenarios, there are in general five parameters that 
describe the two processes under consideration here, namely 
$g_{Z'1}$, $g_{Z'2}$, $g_{Z'1}'$, $g_{Z'2}'$, and $M_{Z'}$, where the coupling 
constants are defined as in Eqs.~(\ref{VecHam}, \ref{VecHamLept}) by 
substiuting $V \to Z^\prime$. There are, of course, many ways to reduce this number.  
In the following, let us assume that $Z'$ couples only to 
left-handed quarks and has SM-like diagonal couplings to leptons, 
\beq
g_{Z'2}=0,\quad 
g_{Z'1}' = \frac{g}{\cos\theta_W} 
\left(-\frac{1}{2} + \sin^2\theta_W \right), \quad
g_{Z'2}' = \frac{g \sin^2\theta_W }{\cos\theta_W} \ \ , 
\eeq
where $g$ is again the SM $SU(2)$ gauge coupling. This 
procedure reduces the number of unknowns to two, $g_{Z'1}$ 
and $M_{Z'}$. Note that for purely vector couplings 
of a $Z'$ to leptons, {\it i.e.} $g_{Z'1}' = g_{Z'2}'$ no 
contributions are generated for  $D^0\to \mu^+\mu^-$ due to 
conservation of vector current.

\vspace{0.2cm} 

\noindent{\it $D^0$-${\bar D}^0$ Mixing}: The contribution 
of the $Z'$ model to mixing is given by Eq.~(\ref{dMV2}),
\begin{eqnarray}\label{zpr}
x^{\rm (Z')}_{\rm D} = {f_D^2 M_D B_D r (m_c, M_{Z'})  
\over 3 \Gamma_D} ~ {g_{Z'1}^2 \over M_{Z'}^2} \ \ . 
\end{eqnarray}
For the very slowly varying RG factor, we have taken 
$r (m_c, M_{Z'}) = 0.71$, which is typical 
of values for a $Z'$ mass in the TeV range.  
From Eq.~(\ref{zpr}), we obtain the bound 
$M_{Z'} / g_{Z'1} \ge 1.7\times 10^6$~GeV.

\vspace{0.2cm} 

\noindent {\it $D^0 \to \mu^+\mu^-$ Decay}: In this model, 
the contribution to the rare decay branching fraction can be 
written in the form, 
\beq\label{GammaZpr}
{\cal B}_{D^0 \to \mu^+\mu^-}^{\rm ({Z'})} = 
\frac{G_F f_D^2 m_\mu^2 M_D}{16\sqrt{2} \pi \Gamma_D}
\sqrt{1-\frac{4 m_\mu^2}{M_D^2}} 
~ {g_{Z'1}^2 \over M_{Z'}^2}\cdot {M_Z^2 \over M_{Z'}^2} \ \ .
\eeq
Besides the $g_{Z'1}^2 / M_{Z'}^2$ dependence which appears in 
the above $D$ mixing relation of Eq.~(\ref{zpr}), there is now 
an additional factor of $M_Z^2/M_{Z'}^2$.  
The bound obtained from Eq.~(\ref{brs}) implies the 
restriction $M_{Z'} / g_{Z'1}^{1/2} \ge 8.7 ~10^2$~GeV, which is weaker 
than the constraint from $D^0$-${\bar D}^0$ mixing.
 
\vspace{0.2cm} 

\noindent {\it Combining the Mixing and Decay Relations}: 
Assuming that $Z'$ 
saturates the observed experimental value for $x_D$, the bound 
obtained from the $D^0\to \mu^+\mu^-$ branching fraction as a 
function of $M_{Z'}$ is  
\beqa
{\cal B}_{D^0\to \mu^+\mu^-} &=&
\frac{3 G_F m_\mu^2 M_Z^2 x_D}{16\sqrt{2} \pi B_D r(m_c,M_{Z'})}
\sqrt{1-\frac{4 m_\ell^2}{M_D^2}}  
{1 \over M_{Z'}^2} \nonumber \\
&\simeq& \ 2.4\times 10^{-10} ~ \left(x_{\rm D}/M_{Z'}^2({\rm TeV})\right)
\ \le \ 2.4\times 10^{-12}/M_{Z'}^2({\rm TeV}) \ \ .
\eeqa
%

\subsection{Family (Horizontal) Symmetries}

The gauge sector in the Standard Model has a large global symmetry which
is broken by the Higgs interaction~\cite{sher}.  
By enlarging the Higgs sector, some
subgroup of this symmetry can be imposed on the full SM lagrangian and
break the symmetry spontaneously.  This family symmetry can be global as well
as gauged~\cite{Monich:1980rr}.  If the new gauge couplings are very weak 
or the gauge boson
masses are large, the difference between a gauged or global symmetry
is rather difficult to distinguish in practice.  In general there would be
FCNC effects from both the gauge and scalar sectors. Here we consider the
gauge contributions.

Consider the group $SU(2)_G$ acting only on the first two 
left-handed families (it may be regarded 
as a subgroup of an $SU(3)_G$, which is broken). Spontaneous 
breaking of $SU(2)_G$ makes the gauge
bosons $G_i$ massive. 
For simplicity we assume that after symmetry breaking the gauge 
boson mass matrix is diagonal to a good approximation in which 
case $G_{i\mu}$ are physical eigenstates and any mixing between
them is neglected.  Leaving further discussion to 
Refs.~\cite{Golowich:2007ka,Burdman:2001tf}, we write down 
the couplings in the fermion mass basis as 
\bea
& & {\cal H}_{hs} = -f \Bigl[
G_{1\mu} 
\Bigl\{
\sin2\theta_d \left(
\overline{d}_L \gamma_\mu d_L - \overline{s}_L \gamma_\mu s_L \right)
+ \sin2\theta_u \left(
\overline{u}_L \gamma_\mu u_L - \overline{c}_L \gamma_\mu c_L \right)
\nonumber \\
&+& \sin2\theta_l \left(
\overline{e}_L \gamma_\mu e_L - \overline{\mu}_L \gamma_\mu \mu_L \right)
+ \cos2\theta_d \left(
\overline{d}_L \gamma_\mu s_L + \overline{s}_L \gamma_\mu d_L \right)
\nonumber \\
&+& \cos2\theta_u \left(
\overline{u}_L \gamma_\mu c_L + \overline{c}_L \gamma_\mu u_L \right)
+ \cos2\theta_l \left(
\overline{e}_L \gamma_\mu \mu_L + \overline{\mu}_L \gamma_\mu e_L \right)
\Bigr\}
\nonumber \\
&+& i G_{2\mu}
\left\{
\left(\overline{s}_L \gamma_\mu d_L - \overline{d}_L \gamma_\mu s_L \right)
+ \left(\overline{c}_L \gamma_\mu u_L - \overline{u}_L \gamma_\mu c_L \right)
+ \left(\overline{\mu}_L \gamma_\mu e_L - \overline{e}_L 
\gamma_\mu \mu_L \right)
\right\}
\nonumber \\
&+& G_{3\mu}
\Bigl\{
\cos2\theta_d \left(
\overline{d}_L \gamma_\mu d_L - \overline{s}_L \gamma_\mu s_L \right)
+ \cos2\theta_u \left(
\overline{u}_L \gamma_\mu u_L - \overline{c}_L \gamma_\mu c_L \right)
\nonumber \\
&+& \cos2\theta_l \left(
\overline{e}_L \gamma_\mu e_L - \overline{\mu}_L \gamma_\mu \mu_L \right)
- \sin2\theta_d \left(
\overline{d}_L \gamma_\mu s_L + \overline{s}_L \gamma_\mu d_L \right)
\nonumber \\
&-& \sin2\theta_u \left(
\overline{u}_L \gamma_\mu c_L + \overline{c}_L \gamma_\mu u_L \right)
- \sin2\theta_l \left(
\overline{e}_L \gamma_\mu \mu_L + \overline{\mu}_L \gamma_\mu e_L \right)
\Bigr\}
\Bigr]\ \ .
\label{bigL}
\eea
Applications of this general interaction yield 
expressions for $D^0$-${\bar D}^0$ mixing, 
\beq
x^{\rm (FS)}_{\rm D} = \frac{2f_D^2 M_DB_D r(m_c,M)}{3\Gamma_D}~ f^2
\left({\cos^2 2\theta_u \over m_1^2} + 
{\sin^2 2\theta_u \over m_3^2}  - {1 \over m_2^2} \right)
   \ \ , 
\label{HSmix}
\eeq
for $D^0 \to \mu^+\mu^-$ decay, 
\beqa
& & {\cal B}_{D^0\to \mu^+\mu^-}^{\rm (FS)} = {M_D f_D^2 m_\mu^2 \over 
64 \pi \Gamma_D} f^4 \left({\sin 2\theta_u \cos 2\theta_\ell \over m_3^2} 
- {\cos 2\theta_u \sin 2\theta_\ell \over m_1^2} \right)^2 \ \ , 
\label{HSmumu}
\eeqa
and for 
$D^0 \to \mu^+e^-$ decay, 
\beqa
& & {\cal B}_{D^0\to \mu^+e^-}^{\rm (FS)} = 
{M_D f_D^2 m_\mu^2 \over 64 \pi \Gamma_D} 
f^4 \left({\cos 2\theta_u \cos 2\theta_\ell \over m_1^2} 
+ {1 \over m_2^2}  + 
{\sin 2\theta_u \sin 2\theta_\ell \over m_3^2} \right)^2 \ \ .
\label{HSmuel}
\eeqa
In Eq.~(\ref{HSmix}) the very-slowly varying RG factor  
$r(m_c,M)$ is set to the scale $M \sim 1$~TeV. 

Precise predictions for the above three processes are not 
immediate due to the large number of NP parameters.  
Different patterns can be obtained depending on the region of 
parameter space:

{\it Case A} [$m_1 = m_3 \ll m_2$ and 
$\theta_u - \theta_\ell = \pi/4$]:

\noindent Here, ${\cal B}_{D^0\to \mu^+e^-}^{\rm (FS)}$ is suppressed 
and a parameter-free prediction for ${\cal B}_{D^0\to \mu^+\mu^-}$ 
in terms of $x_D$ occurs, 
\beqa
{\cal B}_{D^0\to \mu^+\mu^-} = {9 \Gamma_D  m_\mu^2 x_D^2 
\over 256 \pi M_D f_D^2 B_D^2 r(m_c,m_1)^2} 
\simeq 0.7 \times 10^{-14}~ x_D^2 \le  0.7 \times 10^{-18} \ \ .
\label{HSpredict}
\eeqa
Note that here we related the ${\cal B}_{D^0\to \mu^+\mu^-}$ to the {\it square} of $x_D$.

{\it Case B} [$m_1 = m_2 = m_3$ and 
$\theta_u - \theta_\ell = \pi/2$]:

\noindent In this case, the amplitudes for all three processes vanish.

{\it Case C} [$m_1 = m_2 \ll m_3$ and 
$\theta_u - \theta_\ell = \pi/2$]:

\noindent Now, the mixing contribution vanishes but the branching 
fractions for $D^0\to \mu^+\mu^-$ and $D^0\to \mu^+e^-$ are equal, 
although undetermined due to NP parameter dependence.

{\it Case D} [$m_1 = m_3 \gg m_2$]:

\noindent  In this limit, $D^0\to \mu^+\mu^-$ is negligible 
and there is a parameter-free prediction for 
${\cal B}_{D^0\to \mu^+\mu^-}$ in terms of $x_D$, 
but $x_D$ has the wrong sign.

\section{NP Models with loop amplitudes}

Although tree amplitudes represent 
the most obvious situation for producing a correlation between 
mixing and decay, it turns out that loop amplitudes can 
have the same effect.  As is well known~\cite{Inami:1980fz}, 
low energy effective lagrangians continue to provide the most 
useful description.  In the following, we consider three examples 
of NP models with loop amplitudes.

\subsection{Heavy Vector-like Quarks: Q = -1/3 Singlet Quark}

We first consider models with a 
heavy vector-like $Q = -1/3$ singlet quark.  Note that essentially 
identical results hold for a SM fourth quark generation as well, since 
in each case, the fermions will interact with a SM $W^\pm$ gauge boson 
and thus the charged leptons have SM interactions~\cite{q4}.  It is this 
which allows for correlations between 
$D^0$-${\bar D}^0$ mixing and $D^0\to \ell^+\ell^-$ decay.  

For the class of models with $Q = -1/3$ down-type
singlet quarks,  the down quark mass matrix is 
a $4\times 4$ array if there is just one heavy singlet 
(or $6\times 6$ for three heavy singlets as 
in $E_6$ models).  As a consequence, the standard $3\times 3$ CKM 
matrix is no longer unitary.  Moreover, the weak charged current 
will now contain terms that couple up-quarks to the
heavy singlet quarks.  For three heavy singlets, we have 
\beq
{\cal L}_{\rm int}^{(ch)} = {g \over \sqrt{2}}~V_{i\alpha} 
W^\mu {\bar u}_{i,L} \gamma_\mu D_\alpha \ \ , 
\eeq
where $u_{i,L} \equiv (u,c,t)_L$ and $D_\alpha \equiv (D,S,B)$ 
refer to the standard up quark and heavy isosinglet down quark 
sectors.  The $\{ V_{i\alpha} \}$ are elements of a 
$3\times 6$ matrix, which is the product of the $3\times 3$ and
$6\times 6$ unitary matrices that diagonalize the $Q=+2/3$ and 
$Q=-1/3$ quark sectors, respectively.  

\vspace{0.2cm}

\begin{figure}[tbp]
\centerline{
\includegraphics[width=4.5cm,angle=0]{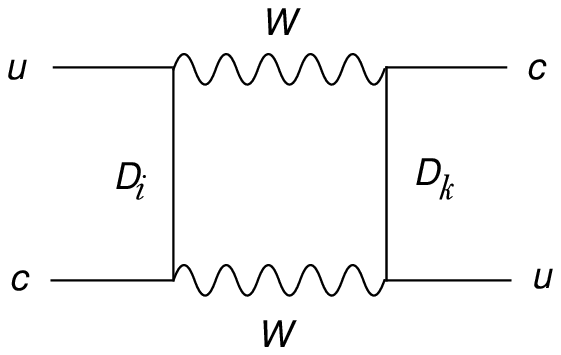}}
\vspace*{0.1cm}
\caption{Box contribution from heavy weak-isosinglet quarks.}
\label{qfig1}
\end{figure}

\noindent {\it $D^0$-${\bar D}^0$ Mixing}: 
The box diagram contribution to 
$\Delta M_{\rm D}$ from these new quarks is displayed in Fig.~\ref{qfig1}. 
Assuming that the contribution of one of the heavy quarks (say the 
$S$ quark, of mass $m_S$) dominates, one can write an expression 
for $x_{\rm D}$~\cite{q4}, 
\beq
|x_{\rm D}^{(-1/3)}| \simeq 
{G_F^2 M_W^2 f_D^2 M_D \over 6 \pi^2 \Gamma_D}  B_D 
\left( V_{cS}^* V_{uS} \right)^2 r(m_c,M_W) |{\overline E}(x_S)|  \ \ , 
\eeq 
where $x_S \equiv (m_S/M_W)^2$.  The Inami-Lin~\cite{Inami:1980fz} 
function ${\overline E}(x_S)$ is defined as 
\beq 
{\overline E}(x_S) \equiv x_S \left[ {1 \over 4} - {9 \over 4 (x_S - 1)} - 
{3 \over 2 (x_S - 1)^2} + {3 x_S^3 \over 2 (x_S - 1)^3}\ln x_S  
\right] \ \ .
\label{il0}
\eeq
For our numerical work, we assume a default 
value of $m_S = 500$~GeV, but express 
our result for variable $m_S$ by noting 
that the functions ${\overline E}(x_S)$ and 
${\bar C}(x_S)$ ({\it cf} Eq.~(\ref{Cbar}) below) are proportional 
to $x_S$ within ten per cent over the mass region $400 
\le m_S({\rm GeV}) \le 700$.  
The light-heavy mixing angles $\left|V^*_{cS} V_{uS}\right|^2$
should go as $1/m_S$ for large $m_S$ to keep the
contribution under control. The current bound on 
$\left|V^*_{cS} V_{uS}\right|^2$ from unitarity of the CKM matrix
is not very stringent, 
$\left|V^*_{cS} V_{uS}\right|^2 < 4\times 10^{-4}$~\cite{PDG}.

In the $E_6$-based model proposed by Bjorken {\it et al}~\cite{q7}, the 
$6\times 6$ mass matrix has an especially simple form. The resulting 
$6\times 6$ mass matrix has a pseudo-orthogonality property which
implies that the $3\times 3$ CKM matrix, although not unitary,
satisfies
\beq
\sum_{i=1}^3 \ \left(V_{\rm CKM}\right)_{bi}^* 
\left(V_{\rm CKM}\right)_{is} = 0\ \ .
\eeq
The analog of this condition in the up quark sector does not hold, and as a
result, there are no new FCNC effects in the down quark sector.  For the CKM
elements participating in
$D^0$-${\bar D}^0$ mixing, the prediction is now  
(recall capital lettering is used to denote the heavy quark) 
\beq
\left| V_{cS}^*V_{uS} \right|^2 = s_2^2 
\left| V_{cs}^*V_{us} \right|^2  \simeq s_2^2 \lambda^2 \ \ ,
\eeq
where $|V_{cs}^*V_{us}| \simeq \lambda \simeq 0.22$ and 
$s_2$ is the (small) mixing parameter describing the mixing between 
the light $s$ quark and the heavy $S$ quark.  Thus, we rewrite 
$|x_{\rm D}^{(-1/3)}|$ in the modified form,  
\beq
|x_{\rm D}^{(-1/3)}| \simeq 
{G_F^2 M_W^2 f_D^2 M_D \over 6 \pi^2 \Gamma_D}  B_D 
s_2^2 \lambda^2 r(m_c,M_W) |{\overline E}(x_S)| \ \ .
\eeq 

\vspace{0.2cm}

\noindent {\it $D^0 \to \mu^+\mu^-$ Decay}: 
For $D^0 \to \mu^+\mu^-$, the effective Lagrangian is given in 
Eq.~(2.2) of Ref.~\cite{Inami:1980fz},
\beqa
& & {\cal L}_{\rm eff} = {G_F^2 M_W^2 \over \pi^2} {\bar C}(x_S) 
\lambda s_2 ~ {\widetilde Q}_1 \ \ , 
\eeqa
where 
\beqa
{\bar C}(x_S) &\equiv&  {x_S \over 4} - {3 x_S \over 4(x_S - 1)}
  - {3 \over 4} 
\left( {x_s \over x_S - 1} \right)^2 \ln x_S \ \ .
\label{Cbar}
\eeqa
In this model, the $D^0 \to \mu^+ \mu^-$ branching fraction becomes 
\beqa
{\cal B}_{D^0 \to \mu^+ \mu^-} = { M_D 
\sqrt{1 - 4 {m_\mu^2 \over M_D^2}}~
\left( G_F M_w \right)^4 \cdot 
 \left(s_2 \lambda f_D m_\mu {\bar C}(x_S)\right)^2
\over 32 \pi^5 \Gamma_{D} } \ \ .
\eeqa

\vspace{0.2cm}

\noindent {\it Combining the Mixing and Decay Relations}: 
If we eliminate $s_2^2$ from the mixing and decay relations, 
we obtain 
\beqa
& & {\cal B}_{D^0 \to \mu^+\mu^-} = {6 \over 32 \pi^3} \cdot 
{x_{\rm D} \sqrt{1 - 4 m_\mu^2/M_D^2}~
\left( m_\mu G_F M_W {\bar C}(x_S)\right)^2 \over 
B_D  r(m_c,M_W) |{\bar E}(x_S)|} \nonumber \\
& & ~ \simeq 1.0 \times 10^{-9} ~x_{\rm D} 
~\left( {m_S \over 500~{\rm GeV}} \right)^2 \ \le \ 
1.0 \times 10^{-11}~\left( {m_S \over 500~{\rm GeV}} \right)^2
 \ \ . \label{x10} 
\eeqa

\subsection{Minimal Supersymmetric Standard Model}

We next consider the Minimal Supersymmetric Standard Model (MSSM) with
unbroken R-parity.  Conservation of R-parity implies that only pairs
of sparticles can be produced or exchanged in loops. We will assume that 
neither squarks nor gluinos are decoupled (direct collider searches
for squark and gluino pair production place the bound
$m_{\tilde q,g}\gsim 330$~GeV~\cite{PDG} in the MSSM with minimal gravity
mediated Supersymmetry breaking), so the MSSM can in principle 
give a dominant contribution to the processes under consideration here.

We will not assume any particular SUSY breaking mechanism, and hence
parameterize all possible soft SUSY-breaking terms.  We work in the so-called
super-CKM basis,  where flavor violation is driven by non-diagonal squark 
mass insertions (see \cite{Golowich:2007ka} for a discussion of this 
mechanism in $\DDbar$ mixing).  In this case, the squark-quark-gluino
couplings are flavor conserving, while the squark propagators are expanded
to include the non-diagonal mass terms.  The $6\times 6$ mass matrix for
the $Q = +2/3$ squarks can be divided into $3\times 3$ sub-matrices,
\beq
\widetilde{M}^2 = 
\left(
\begin{array}{c c}
\widetilde{M}^2_{LL} & \widetilde{M}^2_{LR} \cr
\widetilde{M}_{LR}^{2~T} & \widetilde{M}^2_{RR}
\end{array}
\right) \ \ ,
\eeq
and the mass insertions can be parameterized in a model independent
fashion as
\beq
\left(\delta_{ij}\right)_{MN} = 
{\left(V_M \widetilde{M}^2 V_N^\dagger\right)_{ij} \over 
m_{\tilde q}^2} \ \ .
\label{squarkmat}
\eeq
Here, $i,j$ are flavor indices, $M,N$ refers to the helicity choices 
$LL$, $LR$, $RR$, and $m_{\tilde q}$ represents the average
squark mass. The squark-gluino loops with mass insertions are by far the
largest supersymmetric contribution to $\DDbar$-mixing and can dominate
the transition. The effective 
hamiltonian relevant for this contribution to $\DDbar$-mixing is given by 
\begin{equation}
{\cal H}_{MSSM}^{mix} = {\alpha_s^2\over 2 m^2_{\tilde q}}\sum_{i=1}^8
C_i(m_{\tilde q})Q_i\ \ ,
\end{equation}
where all eight operators contribute in the MSSM.  Evaluating the
Wilson coefficients at the SUSY scale gives, 
\begin{eqnarray}
C_1(m^2_{\tilde q}) & = & {1\over 18}
\left(\delta^u_{12}\right)^2_{LL} [4xf_1(x)+11f_2(x)]\ ,
\nonumber\\
C_2(m^2_{\tilde q})& = & {1\over 18}\left\{
\left(\delta^u_{12}\right)_{LR}\left(\delta^u_{12}\right)_{RL}\, 15f_2(x)
-\left(\delta^u_{12}\right)_{LL}\left(\delta^u_{12}\right)_{RR}
[2xf_1(x)+10f_2(x)]\right\}\ ,\nonumber\\
C_3(m^2_{\tilde q})& = & {1\over 9}\left\{
\left(\delta^u_{12}\right)_{LL}\left(\delta^u_{12}\right)_{RR}
[42xf_1(x)-6f_2(x)]-
\left(\delta^u_{12}\right)_{LR}\left(\delta^u_{12}\right)_{RL}\, 11f_2(x)
\right\}\ ,\nonumber\\
C_4(m^2_{\tilde q})& = & {1\over 18}
\left(\delta^u_{12}\right)^2_{RL} 37xf_1(x)\ ,\nonumber\\
C_5(m^2_{\tilde q})& = & {1\over 24}
\left(\delta^u_{12}\right)^2_{RL}\, xf_1(x)\ ,\\
C_6(m^2_{\tilde q})& = & {1\over 18}
\left(\delta^u_{12}\right)^2_{RR}[4xf_1(x)+11f_2(x)]\ ,\nonumber\\
C_7(m^2_{\tilde q}) & = & {1\over 18}
\left(\delta^u_{12}\right)^2_{LR}37xf_1(x)\ ,\nonumber\\
C_8(m^2_{\tilde q}) & = & {1\over 24}
\left(\delta^u_{12}\right)^2_{LR}\, xf_1(x)\ ,\nonumber
\end{eqnarray}
where $x \equiv m^2_{\tilde g}/m^2_{\tilde q}$, 
with $m_{\tilde g}$ being the mass of the gluino. The equations above
are symmetric under the interchange $L\leftrightarrow R$. 
These contributions to $x_D$ are found to be large \cite{Golowich:2007ka},
and the observation of $D$ mixing constrains the mass insertions to be at
the percent level, or less, for Tev-scale sparticles.

We now examine the squark-gluino contribution to
rare decays, which proceeds through $Z$ penguin diagrams for on-shell
leptons.
The relevant $c \to u \ell^+\ell^-$ Lagrangian is given, for example, 
in Ref.~\cite{Burdman:2001tf}.  Electromagnetic current conservation
forbids the contribution of the photonic penguin diagram for
on-shell leptons in the final state.  In addition, the vector
leptonic operator $\bar\ell\gamma_\mu\ell$ also does not contribute
for on-shell leptons as $p^\mu_D(\bar\ell\gamma_\mu\ell) = (p^\mu_{\ell^+}
+p^\mu_{\ell^-})(\bar\ell\gamma_\mu\ell)=0$.  The effective hamiltonian
is then given by
\begin{equation}
{\cal H}_{MSSM}^{rare} = - \frac{4 G_F}{\sqrt{2}}  \frac{e^2}{16 \pi^2} \left[
c_{10} \ (\overline{\ell} \gamma_\mu \gamma_5 \ell) \ 
(\overline{u}_{L} \gamma^\mu c_L) +
c_{10}^{\prime} \ (\overline{\ell} \gamma_\mu \gamma_5 \ell) \ 
(\overline{u}_{R} \gamma^\mu c_{R})
\right],
\end{equation}
where $c_{10}$ and $c_{10}^{\prime}$ are given by \cite{enrico}
\begin{equation}
%
c_{10}  =  -\frac{1}{9} \frac{\alpha_s}{\alpha} (\delta^u_{22})_{LR} 
(\delta^u_{12})_{RL} P_{032} \,, \quad\quad
c_{10}^\prime  =  -\frac{1}{9} \frac{\alpha_s}{\alpha} (\delta^u_{22})_{RL} 
(\delta^u_{12})_{LR} P_{122} \,.
\end{equation}
$P_{032,122}$ are kinematic loop functions and are defined in the above reference.
The double mass insertion is required to induce a helicity flip in the squark
propagator.  Due to this double mass insertion, this contribution to
$D \to \ell^+\ell^-$ is completely negligible. 
We note that the chargino contribution to the $Z$ 
penguin for $D \to \ell^+\ell^-$ also contains a double mass insertion.  
The leading MSSM contribution to this rare decay is thus most likely mediated by a box
diagram with squark-chargino-sneutrino exchange.  This precludes a relation
to $\DDbar$ mixing. 

Lastly, we note that in contrast to the $B_s$ system \cite{buras}, $D \to \ell^+\ell^-$ 
does not receive a sizable contribution from Higgs boson exchange with large
$\tan\beta$.  This is because in this case, the loop-induced term to the Yukawa couplings
is proportional to $v_d$ ({\it i.e.,} the vev of the Higgs doublet 
that generates masses for the down-type quarks) which is the smaller of the
two vevs and hence does not compensate for the small loop factor.



\subsection{R Parity Violating Supersymmetry}

Finally, we consider Supersymmetry with R-Parity violation (RPV).
We refer the reader to 
Refs.~\cite{Golowich:2007ka,Burdman:2001tf} for 
discussions and earlier references of RPV-SUSY relevant to this
paper. Suffice it to say that the lepton number violating 
RPV-SUSY interactions can be expressed as 
\begin{eqnarray}
& & W_{\lambda'} = \tilde\lambda'_{ijk}\left\{V_{jl}\left[ \tilde\nu^i_L
\bar d^k_Rd^l_L+\tilde d^l_L\bar d^k_R\nu^i_L
+(\tilde d^k_R)^*(\bar\nu^i_L)^cd^l_L\right]  
-\tilde e^i_L\bar d^k_Ru^j_L
-\tilde u^j_L\bar d^k_Ru_L^j-(\tilde d_R^k)^*(\bar e^i_L)^cu_L^j
\right\}, \nonumber \\
\label{lrpar}
\end{eqnarray}
in terms of the coupling parameters 
$\{ \tilde\lambda'_{ijk}\} $.  The generation indices denote the 
correspondences 
$i \Leftrightarrow$ leptons or sleptons, 
$j \Leftrightarrow$ up-type quarks 
and $k \Leftrightarrow$ down-type quarks or squarks.   

\vspace{0.2cm}

\begin{figure} [tb]
\centerline{
\includegraphics[width=9cm,angle=0]{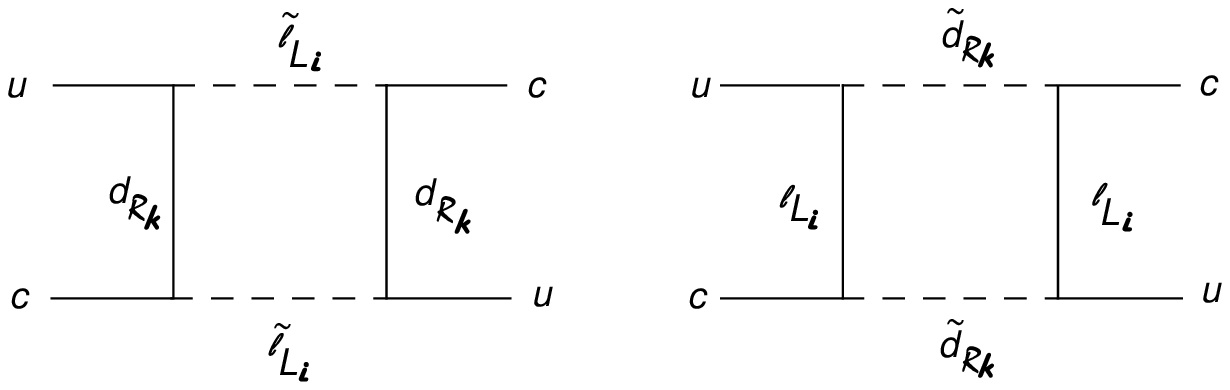}}
\caption{Contributions to $D^0$-$\overline D^0$ mixing from the
$\lambda'$ superpotential terms in supersymmetric
models with R-parity violation.
\label{rpardiag}}
\end{figure}

\noindent {\it $D^0$-${\bar D}^0$ Mixing}: 
As described at the high mass scale by the 
effective hamiltonian 
\begin{equation}
{\cal H}_{R_p}={1\over 128\pi^2}(\tilde\lambda'_{i2k}
\tilde\lambda'_{i1k})^2\left[ {1\over m_{\tilde\ell_{L,i}}^2}
+{1\over m_{\tilde d_{R,k}}^2}\right] Q_1\ \ , 
\end{equation}
this implies constraints on the product of couplings 
$\tilde\lambda'_{i2k}\tilde\lambda'_{i1k}$.  Here, we have
assumed that only one set of the R-parity violating
couplings $\tilde\lambda'_{i2k}\tilde\lambda'_{i1k}$ is large
and dominant.  This is equivalent to saying that, {\it e.g.}, both 
the sleptons and both the down-type quarks being exchanged 
in the first contribution to the box diagram of Fig.~\ref{rpardiag} 
are from the same generation.  In general, this
need not be the case and the coupling factor would then be the
product $\tilde\lambda'_{i2k}\tilde\lambda'_{m1k}
\tilde\lambda'_{m2n}\tilde\lambda'_{i1n}$,
with, {\it e.g.}, the set of $\tilde\ell_{L,i}, d_{R,k}, 
\tilde\ell_{L,m}, d_{R,n}$ 
being exchanged.
Computing the evolution to the charm-quark scale yields 
\begin{equation}
{\cal H}_{R_p}= {1\over 2 m^2_{\tilde d_{R,k}}} C_1(m_c)Q_1\ ,
\end{equation}
with $C_1(m_c)=r(m_c,m_{\tilde q})C_1(m_{\tilde q})$. 
The mixing contribution from the R-parity 
violating $\tilde\lambda'$ terms then implies 
\beqa
& & (\tilde\lambda'_{i2k}\tilde\lambda'_{i1k})^2 
  =  {192\pi^2\Gamma_D m^2_{\tilde d_{R,k}} 
\over (1 + \epsilon) f_D^2 M_D B_D r(m_c,m_{\tilde q})}  
x_{\rm D}  \ \ , 
\eeqa 
where $\epsilon \equiv 
m^2_{\tilde d_{R,k}}/m^2_{\tilde\ell_{L,i}}$.  For definiteness, 
we shall scale the results to the value 
$m_{\tilde d_{R,k}} = 300$~GeV, so that 
\beqa
& & {\tilde\lambda'_{i2k}\tilde\lambda'_{i1k} =  0.0053 \cdot 
{m_{\tilde d_{R,k}}\over 300~{\rm GeV}} \cdot 
\sqrt{2 \over 1 + \epsilon} }  \cdot \sqrt{x_{\rm D} \over 0.01}\ \ .
\eeqa

\vspace{0.2cm}

\noindent {\it $D^0 \to \mu^+\mu^-$ Decay}: 
In RPV-SUSY, the underlying transition for $D^0 \to \mu^+\mu^-$ 
is $c + {\bar u} \to \mu^+ + \mu^-$ via tree-level d-squark exchange.  
The coupling constant dependence for $D^0 \to \ell^+\ell^-$ 
would therefore generally involve 
$\tilde\lambda'_{i2k}\tilde\lambda'_{i1k}$.
For the specific mode $D^0 \to \mu^+\mu^-$, we take $i=2$ to get 
$\tilde\lambda'_{22k}\tilde\lambda'_{21k}$.
The effective hamiltonian of Eq.~(74) in 
Ref.~\cite{Burdman:2001tf}, but with $\ell \to \mu$, reads 
\beqa
& & 
\delta {\cal H}_{\rm eff} = -\frac{\tilde{\lambda}'_{22k}
\tilde{\lambda}'_{21k}} {2 m^2_{\tilde{d}^k_R}}\,
\widetilde Q_1 \ \ . 
\eeqa
This leads to the branching fraction 
\beqa
& & {\cal B}^{\not R_p}_{D^0\to\mu^+\mu^-} = 
\frac{f_D^2\,m_\mu^2\,M_D}{\Gamma_{D}}
\,\left[1-\frac{4m_\mu^2}{M_D^2}\right]^{1/2}
\;\frac{\left({\tilde{\lambda}'_{22k}\tilde{\lambda}'_{21k}}\right)^2}
{128\pi\,m_{\tilde{d}_k}^4} \ \ , 
\label{y2} 
\eeqa
and so the constraint 
\beqa
\left({\tilde{\lambda}'_{22k}\tilde{\lambda}'_{21k}}\right)^2 \le 
{\cal B}_{D^0 \to \mu^+\mu^-} {128\pi\,m_{\tilde{d}_k}^4 \over 
f_D^2 m_\mu^2} ~ {\Gamma_{\rm D} \over  M_{\rm D}}~
\left[ 1 - {4 m_\mu^2 \over M_{\rm D}} \right]^{-1/2} \ \ , 
\label{y3} 
\eeqa
which reads numerically 
\beqa
& & {\tilde{\lambda}'_{22k}\tilde{\lambda}'_{21k}} \le 
0.088~\left( \frac{m_{\tilde{d}_k}}{300~{\rm GeV}}\right)^2 \ \ .
\label{y4}
\eeqa

\vspace{0.2cm} 

\noindent {\it Combining the Mixing and Decay Relations}: 
Note the mixing constraint involves 
${\tilde\lambda'_{i2k}\tilde\lambda'_{i1k}}$, whereas 
the {decay constraint} has 
${\tilde{\lambda}'_{22k}\tilde{\lambda}'_{21k}}$. 
If the $i=2$ case dominates, then we arrive at the prediction 
(here we set $\epsilon = 1$)   
\beqa
{\cal B}^{\not R_p}_{D^0 \to \mu^+\mu^-} &=& {3 \pi m_\mu^2 
\left[ 1 - {4 m_\mu^2/M_{\rm D}} \right]^{1/2} x_D \over 
4 m_{\tilde{d}_k}^2 B_D  r(m_c,m_{\tilde q}) } 
\nonumber \\
&\simeq& \  4.8 \times 10^{-7} ~x_{\rm D} \left( {300~{\rm GeV}\over 
m_{\tilde{d}_k}} \right)^2 \ 
\le \ 4.8\times 10^{-9} \left( {300~{\rm GeV}\over 
m_{\tilde{d}_k}} \right)^2 \ \ . 
\label{relate3}
\eeqa


\section{Conclusions}

\begin{table}[t]
\begin{tabular}{c||c}
\colrule\hline 
Model & ${\cal B}_{D^0 \to \mu^+\mu^-}$ \\ 
\colrule \colrule
Experiment & $\le 1.3 \times 10^{-6}$ \\
Standard Model (SD)& $\sim 10^{-18}$ \\
Standard Model (LD) & $\sim {\rm several} \times 10^{-13}$ \\
$Q=+2/3$ Vectorlike Singlet & $4.3 \times 10^{-11}$ \\
$Q=-1/3$ Vectorlike Singlet & $1 \times10^{-11}~(m_S/500~{\rm GeV})^2$ \\
$Q=-1/3$ Fourth Family & $1 \times 10^{-11}~(m_S/500~{\rm GeV})^2$ \\
$Z'$ Standard Model (LD) & $2.4 \times 10^{-12}/(M_{Z'}{\rm (TeV)})^2$ \\
Family Symmetry & $0.7 \times 10^{-18}$  (Case A)  \\
RPV-SUSY & $~4.8 \times 10^{-9}~(300~{\rm GeV}/m_{\tilde{d}_k})^2$ \\
\colrule\hline
\end{tabular}
\vskip .05in\noindent
\caption{Predictions for $D^0 \to \mu^+\mu^-$ branching fraction for $x_D \sim 1\%$. 
Experimental upper bound is a compilation from~\cite{PDG}.}
\label{tab:corr}
\end{table}

The search for New Physics will in general involve many 
experiments, including the measurement of rare decay branching 
fractions and observation of particle-antiparticle mixing.  
Such experiments are essentially competitors, each seeking 
to be the first to indirectly detect physics beyond the Standard Model.  
At any given point, which measurements 
are more sensitive to New Physics must be determined on a 
case by case basis.  
Our earlier work of Ref.~\cite{Golowich:2007ka} already pointed 
out that the observed $D^0$-${\bar D}^0$ signal imposes 
severe limits for a large number of New Physics models.  
If the $D^0$-${\bar D}^0$ mixing is dominated by one of those 
New Physics contributions, what does it imply for rare decays such as 
$D^0\to \mu^+\mu^-$?   Not only have we been able to answer this
question in several specific scenarios, but we find a striking
correlation in some of the models, wherein the branching fraction for
the decay mode $D^0\to \mu^+\mu^-$ is completely fixed in terms of the
mixing parameter $x_{\rm D}$.

For convenience we have gathered our results in 
Table~\ref{tab:corr}.  For all but one case (Family Symmetry), 
we find the NP branching fraction exceeds the SM branching fraction.  
All the NP branching fractions are, however, well below 
the current experimental bounds of Eq.~(\ref{brs}).  
Anticipating future improvements in sensitivity, 
the first NP model to be constrained will be R-parity 
Violating supersymmetry.  This will require lowering of 
the current bound by a factor of a few hundred.

\acknowledgments
The work of E.G. was supported in part by the U.S.\ National Science
Foundation under Grant PHY--0555304, J.H. was supported by the U.S.
Department of Energy under Contract DE-AC02-76SF00515, 
S.P. was supported by the U.S.\ Department of 
Energy under Contract DE-FG02-04ER41291 and 
A.A.P.~was supported in part by the U.S.\ National Science Foundation under
CAREER Award PHY--0547794, and by the U.S.\ Department of Energy 
under Contract DE-FG02-96ER41005.

\appendix
\section{RG running and mixing matrix elements}

NP contributions are affected by RG-running.  
The relevant $8 \times 8$ anomalous dimension matrix, 
derived to NLO in Ref.~\cite{Ciuchini:1997bw}, is 
applied to LO here (see also Ref.~\cite{Golowich:2007ka})   
to yield the Wilson coefficients, 
\beqa\label{rgrun}
& & C_1 (\mu) = r(\mu,M) ~ C_1 (M) \nonumber \\
& & C_2 (\mu) = r(\mu,M)^{1/2} ~ C_2 (M) \nonumber \\
& & C_3 (\mu) = r(\mu,M)^{1/2}~ {2 C_2 (M)\over 3} +  
r(\mu,M)^{-4} ~\left[ C_3 (M) - {2 C_2 (M)\over 3}\right]
\nonumber \\
& & C_4 (\mu) = r(\mu,M)^{(1+\sqrt{241})/6} ~ \left[ 
\left({1 \over 2} - { 8 \over \sqrt{241}} \right) C_4(M) -
{30 C_5(M)\over \sqrt{241}} \right]    
\nonumber \\
& & \hspace{1.3cm} + r(\mu,M)^{(1-\sqrt{241})/6} ~ \left[ 
\left({1 \over 2} + { 8 \over \sqrt{241}} \right) C_4(M) +
{30 C_5(M)\over  \sqrt{241}} \right]       
\nonumber \\
& & C_5 (\mu) = r(\mu,M)^{(1+\sqrt{241})/6} ~ \left[ 
\left({1 \over 2} + { 8 \over \sqrt{241}} \right) C_5(M) + 
{C_4(M)\over 8 \sqrt{241}} \right]    
\nonumber \\
& & \hspace{1.3cm} + r(\mu,M)^{(1-\sqrt{241})/6} ~ \left[ 
\left({1 \over 2} - { 8 \over \sqrt{241}} \right) C_5(M) -
{C_4(M)\over 8 \sqrt{241}} \right]    
\eeqa
where (presuming that $M > m_t$), 
\beqa
r(\mu,M)&=& \left(\frac{\alpha_s(M)}{\alpha_s(m_t)}\right)^{2/7}
\left(\frac{\alpha_s(m_t)}{\alpha_s(m_b)}\right)^{6/23}
\left(\frac{\alpha_s(m_b)}{\alpha_s(\mu)}\right)^{6/25} \ \ .
\label{wilson}
\eeqa
Regarding the remaining Wilson
coefficients, $C_6$ runs analogous to 
$C_1$ and $C_{7,8}$ run analogous to $C_{4,5}$.  
The presence of operator mixing in Eq.~(\ref{rgrun}) is a 
consequence of the nondiagonal structure of 
the anomalous dimension matrix.

We also need to evaluate the $D^0$-to-${\bar D}^0$ matrix elements of
the eight dimension-six basis operators. In general, this implies
eight non-perturbative parameters that would have to be evaluated
by means of QCD sum rules or on the lattice. We choose those
parameters (denoted by $\{B_i\}$) as follows,
\begin{eqnarray}\label{ME}
& & \begin{array}{l}
\langle Q_1 \rangle = {2 \over 3} f_{\rm D}^2 M_{\rm D}^2 B_1\ , \\
\langle Q_2 \rangle = - {5 \over 6} f_{\rm D}^2 M_{\rm D}^2 B_2 \ , \\
\langle Q_3 \rangle = {7 \over 12} f_{\rm D}^2 M_{\rm D}^2 B_3 \ ,\\
\langle Q_4 \rangle = - {5 \over 12} f_{\rm D}^2 M_{\rm D}^2 B_4 \ ,
\end{array}
\quad \qquad
\begin{array}{l}
\langle Q_5 \rangle = f_{\rm D}^2 M_{\rm D}^2 B_5 \ ,\\
\langle Q_6 \rangle = {2 \over 3} f_{\rm D}^2 M_{\rm D}^2 B_6 \ ,\\
\langle Q_7 \rangle = - {5 \over 12} f_{\rm D}^2 M_{\rm D}^2 B_7 \ ,\\
\langle Q_8 \rangle = f_{\rm D}^2 M_{\rm D}^2 B_8 \ \ ,
\end{array}
\end{eqnarray}
where $\langle Q_i \rangle \equiv \langle {\bar D}^0
| Q_i | D^0 \rangle$, and $f_D$ represents the $D$ meson decay constant.
By and large, the compensatory $B$-factors $\{ B_i \}$ are unknown, 
except in vacuum saturation and in the heavy quark limit; there, one 
has $B_i \to 1$.

Since most of the matrix elements in Eq.~(\ref{ME}) are not known, 
we will need something more manageable in order to obtain numerical
results.  The usual approach to computing matrix elements is to
employ the vacuum
saturation approximation. However, because some of the 
$B$-parameters are known, we introduce a 
`modified vacuum saturation' (MVS), where
all matrix elements in Eq.~(\ref{ME}) are written in terms of (known)
matrix elements of $(V-A)\times (V-A)$ and $(S-P)\times (S+P)$ matrix
elements $B_{\rm D}$ and $B_{\rm D}^{\rm (S)}$, 
%
\begin{eqnarray}\label{ME_MVS}
& & \begin{array}{l}
\langle Q_1 \rangle = \displaystyle{2 \over 3} f_{\rm D}^2 M_{\rm D}^2 B_D \ ,\\
\langle Q_2 \rangle = f_{\rm D}^2 M_{\rm D}^2 B_D \left[ 
- \displaystyle{1 \over 2} - \displaystyle{\eta \over N_c} \right] \ ,\\
\langle Q_3 \rangle = f_{\rm D}^2 M_{\rm D}^2 B_D \left[ 
\displaystyle{1 \over 4 N_c} + \displaystyle{\eta \over 2} \right] \ ,\\
\langle Q_4 \rangle = - \displaystyle{2 N_c - 1 \over 4 N_c} 
f_{\rm D}^2 M_{\rm D}^2 B_D~ \eta \ ,
\end{array}
\qquad \qquad
\begin{array}{l}
\langle Q_5 \rangle = \displaystyle{3 \over N_c} 
f_{\rm D}^2 M_{\rm D}^2 B_D ~ \eta \ , \\
\langle Q_6 \rangle = \langle Q_1 \rangle \ , \\
\langle Q_7 \rangle = \langle Q_4 \rangle \ , \\
\langle Q_8 \rangle = \langle Q_5 \rangle \ \ ,
\end{array}
\end{eqnarray}
where we take $N_c=3$ as the number of colors and define 
\beqa\label{bbar}
& & \eta \equiv {B_{\rm D}^{\rm (S)}\over B_D} 
 \cdot {M_{\rm D}^2 \over m_c^2}  \ \ . 
\eeqa

In our numerical work, we take 
\begin{enumerate}
\item $B_{\rm D}=0.82$, which is the
most recent result from the quenched lattice calculation.  
\item For $\eta$, we use 
$B_{\rm D}^{\rm (S)} \simeq B_{\rm D}$~\cite{Gupta:1996yt} so that 
$\eta \simeq M_{\rm D}^2/m_c^2 \simeq 2$.
\item Regarding the decay constant $f_D$, there is now 
good agreement~\cite{fd} 
between determinations from QCD-lattice simulations 
$f_D^{\rm (latt.)} = 0.207(4)$~GeV and various experiments 
$f_D^{\rm (expt.)} = 0.206(9)$~GeV.  For definiteness, we adopt 
the value $f_D = 0.207$~GeV.  
\end{enumerate}

\end{document}